\documentclass[10pt,journal,a4paper]{IEEEtran}
\usepackage{amssymb}
\usepackage{amsmath}
\usepackage{bm}
\usepackage{comment}
\usepackage{amsthm}
\usepackage{graphicx}
\usepackage{textcomp,gensymb}
\usepackage{subfigure}
\usepackage{float}
\usepackage{cite}
\usepackage{enumerate}
\usepackage{multirow}
\usepackage{amsfonts}
\usepackage{url}
\usepackage{diagbox}
\usepackage{color, soul}
\usepackage{amsthm}
\usepackage{type1cm}

\newcommand{\titlefont}{\fontsize{16.5pt}{24pt}\selectfont}

\newenvironment{proofblack}{{\it Proof:\it}}{\hfill $\blacksquare$\par}
\newtheorem{remark}{\bf~~Remark}

\newtheorem{theorem}{\bf~~Theorem}

\begin{document}
\title{{\titlefont Directivity-Aware Degrees of Freedom Analysis for Extremely \\ Large-Scale MIMO}}

\author{
\IEEEauthorblockN{
    {Shaohua Yue}, \IEEEmembership{Graduate Student Member, IEEE},
	{Liang Liu}, \IEEEmembership{Senior Member, IEEE},
	{and Boya Di}, \IEEEmembership{Member, IEEE}}
	\vspace{-7mm}
    
    \thanks{Shaohua Yue and Boya Di are with the State Key Laboratory of Advanced Optical Communication Systems and Networks, School of Electronics, Peking University, Beijing 100871, China. (email: \{yueshaohua; boya.di\}@pku.edu.cn).} 
    \thanks{Liang Liu is with the Department of Electronic and Information Engineering, The Hong Kong Polytechnic University, Hong Kong, SAR, China. (e-mail: liang-eie.liu@polyu.edu.hk).}
}

\maketitle

\begin{abstract}
Extremely large-scale multiple-input multiple-output (XL-MIMO) communications, enabled by numerous antenna elements integrated into large antenna surfaces, can provide increased effective degree of freedom (EDoF) to achieve high diversity gain. However, it remains an open problem that how the EDoF is influenced by the directional radiation pattern of antenna elements. 
In this work, empowered by the wavenumber-domain channel representation, we analyze the EDoF in a general case where the directivity of antennas, determined by the antenna structure and element spacing, is considered. 
Specifically, we first reveal the uneven distribution of directivity-aware wavenumber-domain coupling coefficients, i.e., channel gain towards different directions, in the isotropic Rayleigh fading channel. 
EDoF is then calculated based on such distribution of coupling coefficients. 
A numerical method is also provided to obtain coupling coefficients via full-wave EM simulations. 
Due to the influence of antenna directivity, how EDoF and ergodic channel capacity vary with the element spacing are explored via simulations for different antenna types.
\end{abstract}


\section{Introduction}\label{sec::intro}

{\color{blue}Extremely large-scale multiple-input multiple-output (XL-MIMO) is an emerging technology for 6G communications, where enormous antenna elements are integrated into the antenna surface for spectral efficiency improvement\cite{xlmimo,xlmimo2}.} {\color{black}Potential energy-efficient implementation methods for XL-MIMO include reconfigurable holographic surfaces\cite{rhs} and reconfigurable intelligent surfaces (RIS)\cite{RIS}}. 
Benefiting from the enlarged antenna aperture, XL-MIMO can achieve an increased effective degree of freedom (EDoF), which is defined as the number of dominant eigenvalues of the XL-MIMO channel correlation matrix\cite{landaudof}. EDoF corresponds to the number of orthogonal communication modes for effective data transmission{\color{black}, and thus,} is considered as an essential figure of merit to evaluate the spatial multiplexing of the MIMO system\cite{edofimportance}.

In the literature, most existing works on MIMO EDoF analysis generally assume ideal antenna arrays with hypothetical antenna elements and quasi-continuous aperture.  In \cite{dofcount}, line-of-sight communications involving large intelligent surfaces are analyzed using the Helmholtz wave equation to derive the {\color{black}EDoF of MIMO channels}.  A wavenumber-domain Fourier series expansion is provided in \cite{wndchannel} to model the Rayleigh fading channel of holographic MIMO and utilized to analyze the ergodic channel capacity and EDoF. The EDoF and channel capacity for MIMO with uniform circular arrays are studied in \cite{oam} to show the advantage of orbital angular momentum-based communication.

{\color{black}However, the above existing works have not considered the influence of antenna elements' directivity on the EDoF of the MIMO system. 
Since the directional radiation pattern of antenna elements has an effect on the channel responses, different types of antennas that exhibit different directivity features naturally lead to various EDoF and channel capacity performances.
 Moreover,} for XL-MIMO with quasi-continuous antenna arrays, where the element spacing is smaller than half of the wavelength, a strong mutual coupling effect between antenna elements arises\cite{antenna}. Such an effect changes the directivity, thereby affecting the EDoF. Therefore, the impact of antenna directivity should be emphasized in this case.

To address the above issue, in this letter, considering the existence of antenna directivity, we analyze how the EDoF and channel capacity of XL-MIMO are influenced by the antenna type and element spacing. 
{\color{black}The wavenumber-domain channel representation is adopted, so that channel gain towards different directions is analyzed efficiently and explicitly using wavenumber-domain coupling coefficients between the transceiver sides.}
We reveal the uneven distribution of coupling coefficients even for the isotropic Rayleigh fading channel caused by antenna directivity. 
Given such distribution information, a method to obtain the EDoF via only wavenumber-domain statistical information is proposed. 
Coupling coefficients of three representative antenna arrays, including patch, dipole, and RIS, are then obtained to calculate EDoF and channel capacity. 
How EDoF and channel capacity of different antenna arrays change with element spacing is also discussed. 

The rest of this letter is organized as follows. In section II, the signal model and the wavenumber domain channel representation of XL-MIMO are described. {\color{black}In Section III, we present the analysis on wavenumber-domain coupling coefficient, which is utilized to obtain the EDoF.} Numerical results of coupling coefficients, EDoF and ergodic channel capacity for three specific types of antennas are given in Section IV. Finally, conclusions are drawn in Section V.
\section{System Model}\label{sec::sys_mod}

\subsection{Scenario Description}

\begin{figure}[t]
	\centerline{\includegraphics[width=7cm,height=4cm]{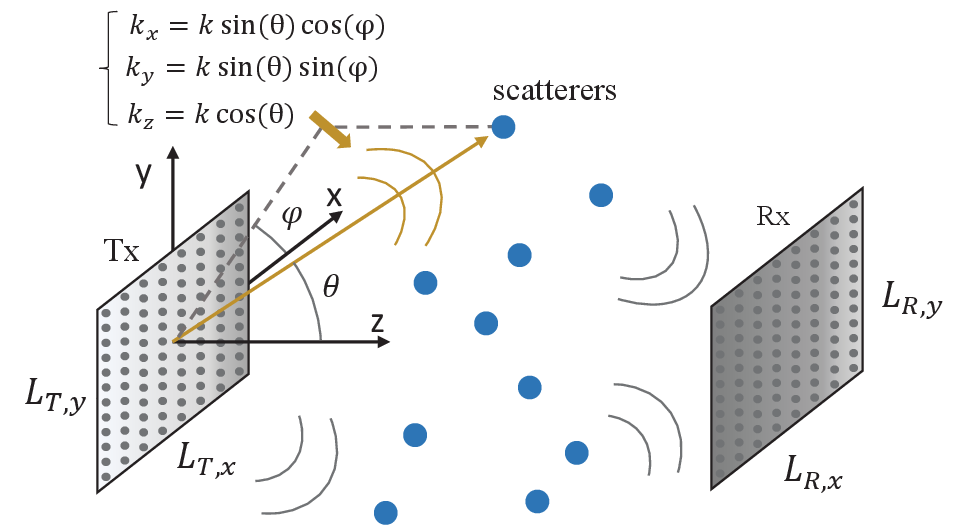}}
	\vspace{-3mm}
	\caption{An XL-MIMO communication system in isotropic scattering Rayleigh channel.}
	\vspace{-4mm}
	\label{fig::scenario}
\end{figure}

As shown in Fig.~\ref{fig::scenario}, we consider an XL-MIMO communication system, where a transmit antenna array with $N_T$ antenna elements communicates with a receive antenna array with $N_R$ antenna elements. A Cartesian coordinate system is introduced, where the $x$-axis and $y$-axis are aligned with the transmit antenna array. The location of the center of the transmit antenna array is set to be the origin, as depicted in Fig.~\ref{fig::scenario}.
The sizes of the transmit and receive antenna arrays 
are denoted as $(L_{T,x}, L_{T,y})$ and $(L_{R,x}, L_{R,y})$, respectively. 

We assume an isotropic scattering Rayleigh fading channel, where numerous scatterers are randomly and evenly distributed in space to generate multipaths. 
The signal model of the XL-MIMO communication system is 
\begin{equation}
\bf{y} = \bf{H}\bf{x} + \bf{n},
\end{equation}
where $\mathbf{x} \in \mathbb{C}^{N_{T} \times 1}\sim \mathcal{CN}(0,\mathbf{Q}), \mathbf{y}\in \mathbb{C}^{N_{R} \times 1}$, $\mathbf{n}\in \mathbb{C}^{N_{R} \times 1}\sim \mathcal{CN}(0,\mu^2)$, and $\mathbf{H} \in \mathbb{C}^{N_{R} \times N_T}$ denote the transmitted signal, the received signal, {\color{black}the zero-mean AWGN noise}, and the random Rayleigh fading channel, respectively. The ergodic channel capacity is expressed as
\begin{equation}{\color{black}
C = \mathbb{E}\{\sum_{i=1}^{\eta}log_2\left(1+\tau_i\left(\mathbf{H}\mathbf{Q}\mathbf{H}^H\right)/\mu^2\right)\}},\label{equ::channel_capacity}
\end{equation}
where 
$\tau_i(\cdot)$ denotes the $i$-th largest eigenvalue of the specified matrix, and $\eta = {\rm Rank}(\mathbf{H}\mathbf{Q}\mathbf{H}^H)$ denotes the degree of freedom (DoF), which indicates the number of positive eigenvalues of the correlation matrix $\mathbf{H}\mathbf{Q}\mathbf{H}^H$. 

\vspace{-4mm}
\subsection{Wavenumber Domain Channel Representation}
{\color{black}It is proved that a spatial stationary and complex Gaussian distribution Rayleigh fading channel can be represented by a finite number of distinguishable planar EM waves, which is the wavenumber-domain channel representation}\cite{wndchannel}.
{\color{black}For a planar wave of wavelength $\lambda$, its wavenumber is denoted as $k = \frac{2\pi}{\lambda}$. The wavenumber is decomposed in the Cartesian coordinate in the form of wavevector $(k_x,k_y,k_z)$ according to the propagation direction of the planar wave, which satisfies $k_x^2+k_y^2+k_z^2 = k^2$\cite{antenna} and the waveumber domain is defined as $\{(k_x,k_y)\in \mathbb{R}^2: k_x^2+k_y^2 \leq k^2\}$.
The wavevectors of planar waves that constitute the wavenumber-domain channel representation are uniformly sampled on the wavenumber domain as at the transmitter and receiver side\cite{wndchannel}, respectively, as 
\begin{equation}
	(k_x,k_y) = (\frac{2\pi m_x}{L_{T,x}},\frac{2\pi m_y}{L_{T,y}}), (\kappa_x,\kappa_y) = (\frac{2\pi l_x}{L_{R,x}},\frac{2\pi l_y}{L_{R,y}}),\label{equ::wvddefinition}
\end{equation}
where $(m_x,m_y)$ and $(l_x,l_y)$ are given as 
\begin{align}
	&\bm{\varepsilon}_T=\{(m_x,m_y)\in \mathbb{Z}^2:(\frac{2\pi m_x}{L_{T,x}})^2+(\frac{2\pi m_y}{L_{T,y}})^2\leq k^2\},\label{equ::wvtx}\\
	&\bm{\varepsilon}_R=\{(l_x,l_y)\in \mathbb{Z}^2:(\frac{2\pi l_x}{L_{R,x}})^2+(\frac{2\pi l_y}{L_{R,y}})^2\leq k^2\}.\label{equ::wvrx}
\end{align}
Hence, the spatial-domain channel $\mathbf{H}$ can be transformed into the wavenumber domain as
\begin{equation}
\mathbf{H} = \sqrt{(N_TN_R)}\mathbf{\Phi}_R \mathbf{H}_a \mathbf{\Phi}_T^H,
\end{equation}
where the wavenumber-domain channel response $\mathbf{H}_a \in \mathbb{C}^{n_{R} \times n_{T}}$ represents the coupling between each pair of planar waves at the transmitter side and the receiver side defined in (\ref{equ::wvddefinition}). {\color{blue}Terms $n_T$ and $n_R$ are} the cardinality of $\bm{\varepsilon}_T$ and $\bm{\varepsilon}_R$, respectively. Because of the spatial stationarity and Gaussian distribution, each element of $\mathbf{H}_a$ satisfies
	\begin{equation}
		\mathbf{H}_a(m_x,m_y;l_x,l_y)  \sim \mathcal{CN}(0,\sigma^2(m_x,m_y;l_x,l_y)),\label{equ::cc}
	\end{equation}
where $\sigma^2(m_x,m_y;l_x,l_y)$ is the coupling coefficient between the pair of planar waves. Terms $\mathbf{\Phi}_T\in \mathbb{C}^{N_{T} \times n_{T}}$ and $\mathbf{\Phi}_R\in \mathbb{C}^{N_{R} \times n_{R}}$ are transform matrices that map the spatial domain to the wavenumber domain and are denoted as
\begin{align}
	&\mathbf{\Phi}_T(i;m_x,m_y)=\frac{1}{\sqrt{N_T}}e^{-{\rm j}(k_{x}s_{x,i}+k_{y}s_{y,i}+k_{z}s_{z,i})},\\
	&\mathbf{\Phi}_R(j;l_x,l_y)=\frac{1}{\sqrt{N_R}}e^{-{\rm j}(\kappa_{x}r_{x,j}+\kappa_{y}r_{y,j}+\kappa_{z}r_{z,j})},
\end{align}
where $k_{x}, k_{y}, \kappa_{x}, \kappa_{y}$ are defined in  (\ref{equ::wvddefinition}). The term $k_{z}= \sqrt{k^2-k_{x}^2-k_{y}^2}$ and $\kappa_{z}= \sqrt{k^2-\kappa_{x}^2-\kappa_{y}^2}$. $(s_{x,i}, s_{y,i},s_{z,i})$ and $(r_{x,j}, r_{y,j},r_{z,j})$ denote the position of the $i$-th element of the transmit array and the $j$-th element of the receive array, respectively.
The wavenumber-domain channel representation $\mathbf{H}_a$ extract channel responses towards different directions via Fourier transform mathematically, which is exploited in the analysis in Section \ref{sec::wndof}. Besides, because $n_T$ and $n_R$ is irrelevant to $N_T$ and $N_R$, the dimension of $\mathbf{H}_a$ depends on the antenna aperture, rather than the number of antenna elements\footnote{In contrast, spatial-domain channel representation $\mathbf{H}$ can lead to a huge matrix dimensions brought by the decreased element spacing.}.}

Note that coupling coefficients $\sigma^2$, as the variance of the wavenumber-domain channel representation, indicate the channel gain distribution in the wavenumber domain and can be used to derive ergodic channel capacity of the XL-MIMO. 
{\color{blue}For the isotropic scattering Rayleigh Fading channel, the joint angle distribution of multipath at the transmit side and the receive side are separate. Therefore, the average angular transfer power between the transmit and receive sides is constant \cite{wndchannel}. In this way, the coupling coefficient is separable and can be decomposed as the multiplication of the transmitter coupling coefficient $\sigma^2_T(m_x,m_y)$ and the receiver coupling coefficient $\sigma^2_R(l_x,l_y)$, i.e., $\sigma^2(m_x,m_y;l_x,l_y) = \sigma^2_T(m_x,m_y)\sigma^2_R(l_x,l_y)$.}
 Thus, the wavenumber-domain channel can be expressed as \cite{wndchannel} $\mathbf{H}_a = {\rm diag}(\bm{\sigma}_R)\mathbf{H}_w{\rm diag}(\bm{\sigma}_T)$,  where each element of $\mathbf{H}_w \sim \mathcal{CN}(0, 1)$. 
If the channel is unknown to the transmitter, the transmit power is allocated uniformly to each data stream, i.e., $\mathbf{Q} = \frac{1}{n_T}\mathbf{\Phi}_T\mathbf{I}_{n_T}\mathbf{\Phi}_T^H$. Hence, the ergodic channel capacity in (\ref{equ::channel_capacity}) is rewritten as
{\color{blue}	\begin{align}
		C = \mathbb{E}\{\sum_{i=1}^{\eta_{u}} log_2(&1+\frac{ N_TN_R}{n_T\mu^2}\tau_i(\mathbf{H}_w{\rm diag}(\bm{\sigma}_T\odot \bm{\sigma}_T)\nonumber \\
		&\times \mathbf{H}_w^H{\rm diag}(\bm{\sigma}_R\odot \bm{\sigma}_R)))\},\label{equ::channelcapwnd}
	\end{align}}
where $\odot$ denotes the inner product, and $(\bm{\sigma}_T\odot \bm{\sigma}_T) \in \mathbb{C}^{n_T \times 1}$ and $(\bm{\sigma}_R\odot \bm{\sigma}_R) \in \mathbb{C}^{n_R \times 1}$ collect all the coupling coefficients $\sigma^2_T(m_x,m_y)$ and $\sigma^2_R(l_x,l_y)$ of the transmitter and the receiver, respectively. By considering the cardinality of $\bm{\varepsilon}_T$ and $\bm{\varepsilon}_R$ defined in (\ref{equ::wvtx}) and (\ref{equ::wvrx}), $\eta_{u}$ is given by
\begin{equation}
	\eta_{u} = \min\{\lfloor\frac{\pi L_{T,x}L_{T,y}}{\lambda^2} \rfloor,\lfloor\frac{\pi L_{R,x}L_{R,y}}{\lambda^2} \rfloor\},\label{equ::dofub}
\end{equation}
 {\color{blue}which is the upper bound of EDoF\cite{wndchannel}. Based on (\ref{equ::channelcapwnd}), both the number of antenna elements and the aperture size of antenna arrays have an effect on the channel capacity, which is consistent with \cite{channelcapacity}.}

\section{Wavenumber-domain Coupling Coefficients} \label{sec::wndof}
In this section, we first define the directivity-aware coupling coefficients $\bm{\sigma}_R, \bm{\sigma}_T$ and analyze the EDoF given the uneven distribution of coupling coefficients in the wavenumber domain. Next, we elaborate on how to obtain coupling coefficients through a numerical method.

\vspace{-4mm}
\subsection{Influence of Directivity on Coupling Coefficients}
Since the antenna's radiation pattern is directional in reality, antenna arrays have different EM responses towards different propagation directions. Consequently, the existence of antenna directivity should be reflected in the calculation of coupling coefficients in (\ref{equ::cc}). {\color{blue}Here, the calculation of directivity-aware coupling coefficients\footnote{\color{blue}The analysis also applies to coupling coefficients at the receiver side. }  ${\sigma}_T^2(m_x,m_y)$
is given as
\begin{equation}
\sigma_T^2(m_x,m_y)=\iint_{\Omega_T(m_x,m_y)}p(\theta_T,\phi_T)G(\theta_T,\phi_T)d\theta_Td\phi_T,\label{equ::calce}
\end{equation}
where $p(\theta_T,\phi_T)$ denotes the angle distribution of the multipath and $G(\theta_T,\phi_T)$ denotes the radiation pattern of the antenna element. We set $\overline{k_x} = \sin(\theta_T)\cos(\phi_T),  \overline{k_y} = \sin(\theta_T)\sin(\phi_T)$, so that the integral region $\Omega_T(m_x,m_y)$ is transformed from the angular domain to the wavenumber domain normalized by the wavenumber $k$ and is given as 
\begin{align}
&K_T(m_x,m_y) = \{(\overline{k_x},\overline{k_y})\in \mathbb{R}^2|\frac{m_x\lambda}{L_{T,x}}\leq \overline{k_x} \leq\frac{(m_x+1)\lambda}{L_{T,x}}, \nonumber \\
&\frac{m_y\lambda}{L_{T,y}}\leq \overline{k_y} \leq\frac{(m_y+1)\lambda}{L_{T,y}},\overline{k_x}^2+\overline{k_y}^2\leq 1\}.\label{equ::intregion}
\end{align}}

Below we show an example of how the antenna directivity affects the coupling coefficients. The antenna directivity is given as
\begin{equation}
G(\theta_T,\phi_T) = \cos^m(\theta_T), 0 \leq \theta_T \leq \pi/2, 0 \leq \phi_T \leq 2\pi,\label{equ::antennadirectivity}
\end{equation}
 where $m \geq 0$ is the antenna directivity coefficient.
This case represents a sufficiently general antenna radiation pattern model \cite{antenna}. 
\vspace{-2mm}

\begin{figure}[th]
	\centerline{\includegraphics[width=9cm,height=3cm]{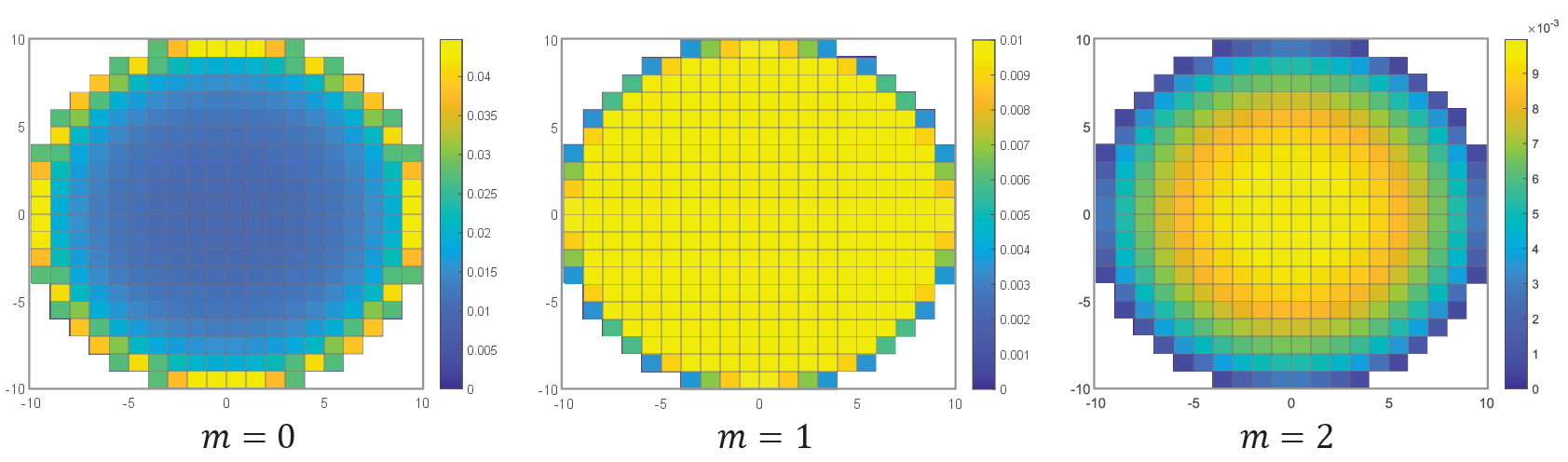}}
	\vspace{-2mm}
	\caption{Distribution of coupling coefficients for different cases of $m$.}
	\label{fig::dccm}
\end{figure}
\begin{theorem}\label{theo::cc}
For antenna elements that satisfy (\ref{equ::antennadirectivity}) in the isotropic scattering Rayleigh fading channel, coupling coefficient $\sigma_T^2(m_x,m_y)$ is given as
\begin{equation}\color{blue}
\sigma_T^2(m_x,m_y) = \frac{1}{2\pi}\iint_{K_T(m_x,m_y)}{(1-\overline{k}_x^2-\overline{k}_y^2)}^{\frac{m-1}{2}}d\overline{k_x}d\overline{k_y}.
\label{equ::calculation_cc}\end{equation}
\end{theorem}

\begin{proofblack}
For the isotropic scattering environment in the half-space above the antenna array, the angle distribution function is given as \cite{dofst}
\begin{equation}
p(\theta_T,\phi_T)=\frac{\sin(\theta_T)}{2\pi}, 0\leq \theta_T \leq\pi/2,0 \leq \phi_T \leq 2\pi.\label{equ::adf}
\end{equation}

The integral variables in (\ref{equ::calce}) are changed from $d\theta_Td\phi_T$ to $d\overline{k_x}d\overline{k_y}$ based on the Jacobian matrix, which is
\begin{equation}
	|\frac{\partial(\theta_T,\phi_T)}{\partial(\overline{k_x},\overline{k_y})}| = {(\overline{k_x}^2+\overline{k_y}^2)}^{-0.5}{(1-\overline{k_x}^2-\overline{k_y}^2)}^{-0.5}.\label{equ::jacobian}
\end{equation}
The angle distribution function and the antenna radiation pattern in the wavenumber domain are given as
\begin{equation}
	p(\overline{k_x},\overline{k_y}) = \frac{\sqrt{\overline{k_x}^2+\overline{k_y}^2}}{2\pi},G(\overline{k_x},\overline{k_y}) = {(1-\overline{k_x}^2-\overline{k_y}^2)}^{\frac{m}{2}}.\label{equ::pginwn}
\end{equation}
The proof is completed by substituting (\ref{equ::pginwn}) into the equation of coupling coefficients calculation (\ref{equ::calce}). 
\end{proofblack}

\begin{remark}\label{rmk::1}
\rm 
According to Theorem \ref{theo::cc}, we show the effect of antenna directivity in Fig. \ref{fig::dccm}. When $m = 0$, i.e., hypothetical antenna elements with $G=1$ for all directions are considered, we have a bowl-shape distribution of coupling coefficients. The case when $m= 1$ indicating a uniform coupling coefficients in the wavenumber domain,  $\sigma_T^2(m_x,m_y) = \frac{\lambda^2}{2\pi L_{T,x}L_{T,y}}$ as long as the integral region in (\ref{equ::intregion}) does not intersect with the circle defined by $\overline{k_x}^2+\overline{k_y}^2= 1$. When $m > 1$, the coupling coefficient reaches the maximum at the center of the wavenumber domain.
\end{remark}

According to the equivalent circuit theory\cite{equcircuit}, the change in element spacing $d$ results in modified RLC equivalent circuit and consequently affects the radiation pattern $G$. 
Therefore, it can be inferred that antenna arrays with various element spacing can give rise to different coupling coefficients and wavenumber-domain channel responses.

\subsection{EDoF Analysis via Coupling Coefficients}

Note that coupling coefficients are evenly distributed in the wavenumber domain
only when antenna elements satisfying $G(\theta,\phi) = \cos(\theta)$ in the isotropic scattering Rayleigh fading channel. 
Since the coupling coefficient is the {channel gain} from the transmitter to receiver at a certain propagation direction\cite{wndchannel}, the uneven distribution of coupling coefficients indicates {high channel gain} only at a restricted range of directions in space, i.e., a subarea of the wavenumber domain, leading to a limited number of available communication modes. 

Therefore, considering the general case of an uneven distribution of coupling coefficients, the calculation of EDoF, which is denoted as ${\eta}_{e}$, based on the statistical information of the wavenumber domain channel is given as
\begin{align}
 &\frac{\sum_{i=1}^{{\eta}_{e,R}}\sigma_{R,i}^2}{\sum_{i=1}^{n_R}\sigma_{R,i}^2} \geq \gamma,\frac{\sum_{i=1}^{{\eta}_{e,T}}\sigma_{T,i}^2}{\sum_{i=1}^{n_T}\sigma_{T,i}^2} \geq \gamma,  \label{equ::wndof}\\
 &{\eta}_{e} = \min\{{\eta}_{e,R},{\eta}_{e,T}\},\label{equ::wndof2}
\end{align}
where $\gamma < 1$ is a threshold to ensure that all dominant coupling coefficients are included in the calculation of $\eta_{e}$. 

The upper bound  for the EDoF, i.e., $\eta_{u}$ given in (\ref{equ::dofub}), is the total number of wavenumber-domain coupling coefficients. 
Differently, $\eta_{e}$ is the number of large coupling coefficients. 
By introducing such an EDoF definition, the coupling coefficients with small values are excluded since they represent electromagnetic (EM) responses with low channel gains in the wavenumber domain unable to transmit information at a high data rate. 
Besides, a similar result of channel capacity can be obtained by replacing $\eta_{u}$ in (\ref{equ::channelcapwnd}) with $\eta_{e}$, further showing that EDoF matches the number of orthogonal communication modes for effective data transmission, which will be demonstrated in Section \ref{sec::sim}.


\subsection{Numerical method to Obtain Coupling Coefficients}
In reality, the analytical expression of the antenna radiation pattern is intractable because of the complicated antenna design. In this way, $(\bm{\sigma}_T \odot \bm{\sigma}_T)$  defined in (\ref{equ::calce}) cannot be derived from directly calculating the integral. Therefore, we provide a EM-simulation-based method to obtain coupling coefficients (EMCC), which primarily includes three steps. First, we obtain the numerical result of antenna directivity via full-wave simulation. Second, we simulate the random spatial-domain channel with the antenna directivity and obtain the wavenumber-domain channel using the LS method. Third, we perform the normal distribution fitting with the simulated wavenumber-domain channel to obtain coupling coefficients.
\begin{enumerate}[i.]
\item \textbf{Numerical Radiation Pattern Simulation}. Obtain the numerical result of the antenna realized gain using full-wave simulation software such as CST, where the simulation boundary is set as unit cell. 
In this way, the mutual coupling effect between elements is simulated\cite{antenna}.
\item \textbf{Wavenumber-domain Channel Simulation.} 
{\color{blue}The channel of the antenna element at location $(x,y)$ on the transmit array is given as\footnote{\color{blue}We mainly consider the small-scale fading effect in this paper for simplicity. The large-scale fading does not affect the pattern of coupling coefficient distribution and EDoF of the XL-MIMO system.}
\begin{equation}
h(x,y) = \frac{1}{\sqrt{S}}\sum_{s=1}^S\sqrt{G(\theta_{T,s},\phi_{T,s})}e^{i(k_{x,s}x+k_{y,s}y)},
\end{equation}
where $S$ is the number of multipaths, $(\theta_{T,s},\phi_{T,s})$ is the propagation direction of the $s$-th multipath, and $(k_{x,s},k_{y,s})$ are the wavevector of the $s$-th multipath. $(\theta_{T,s},\phi_{T,s})$ is sampled from the angle distribution function $p(\theta_T,\phi_T)$ in (\ref{equ::adf}).} 

Note that the simulated spatial domain channel can be approximated with a finite number of wavenumber-domain Fourier series\cite{wndchannel}, which is denoted as
\begin{equation}\color{blue}
    h(x,y) \approx \mathop{\sum\sum}_{(m_x,m_y)\in \bm{\varepsilon}_T}h_a(m_x,m_y)e^{{\rm j}2\pi(\frac{xm_x}{L_{T,x}}+\frac{ym_y}{L_{T,y}})}, \label{equ::wndchannel}
\end{equation}
where $h_a(m_x,m_y) \sim \mathcal{CN}(0,2\sigma_T^2(m_x,m_y))$. Therefore, (\ref{equ::wndchannel}) can be transformed into a linear equation system $\mathbf{E}\mathbf{\tilde{h}} = \mathbf{h}$, where $\mathbf{h}\in \mathbb{C}^{N_{T} \times 1}$ collects the spatial domain channel of each antenna element, $\mathbf{E}\in \mathbb{C}^{N_{T} \times n_T}$ collects the coefficients {\color{blue}$e^{{\rm j}2\pi(\frac{xm_x}{L_{T,x}}+\frac{ym_y}{L_{T,y}})}$} for each antenna element's position and each $(m_x,m_y)\in \bm{\varepsilon}_T$. Note that $N_{T}>n_T$ for $d\leq \lambda/2$, $\mathbf{\tilde{h}}\in \mathbb{C}^{n_{T} \times 1}$ collecting all $\tilde{h}_{m_x,m_y}$ is given with the LS method as
\begin{equation}
    \mathbf{\tilde{h}} = (\mathbf{E}^{H}\mathbf{E})^{-1}\mathbf{E}^{H}\mathbf{h}.
\end{equation}

\item \textbf{Distribution Fitting}.
Step 2 is performed $I$ times to obtain a series of wavenumber-domain channels $\{\mathbf{\tilde{h}}_1,\mathbf{\tilde{h}}_2,...\mathbf{\tilde{h}}_I\}$. The normal distribution fitting is then performed to obtain the variance $2\sigma_T^2(m_x,m_y)$, i.e., the coupling coefficients.
\end{enumerate}

Based on the obtained coupling coefficients $\sigma_T^2(m_x,m_y)$ and $\sigma_R^2(l_x,l_y)$, we can calculate the ergodic channel capacity via (\ref{equ::channelcapwnd}) and the EDoF via (\ref{equ::wndof}) and (\ref{equ::wndof2}).
\section{Numerical Results}\label{sec::sim}
In this section, we first show the effectiveness of proposed EMCC in obtaining coupling coefficients. Next, we discuss the effect of different antenna types and element spacing on coupling coefficients, EDoF, and ergodic channel capacity. Three types of antennas arrays, which are a 26GHz RIS\cite{RIS}\footnote{\color{blue}Since the analysis of coupling coefficients focuses on the channel between the antenna and scatterers in the wireless environment, the analysis applies to both passive RIS and active RIS\cite{activeris}. }, a 2.4GHz dipole antenna array, and a 3GHz patch antenna array, are used as examples for simulation. An antenna array with elements satisfying hypothetical radiation pattern $G(\theta,\phi)=1,0\leq \theta \leq \pi/2,0 \leq \phi \leq 2\pi$ is also considered for comparison. {\color{black}Element spacing larger than half wavelength is not considered to avoid the grating lobe effect\cite{antenna}.}
The size of the transmit and receive arrays are both set as $10\lambda \times 10\lambda$.

\begin{figure}[t]
\centerline{\includegraphics[width=5.6cm,height=4cm]{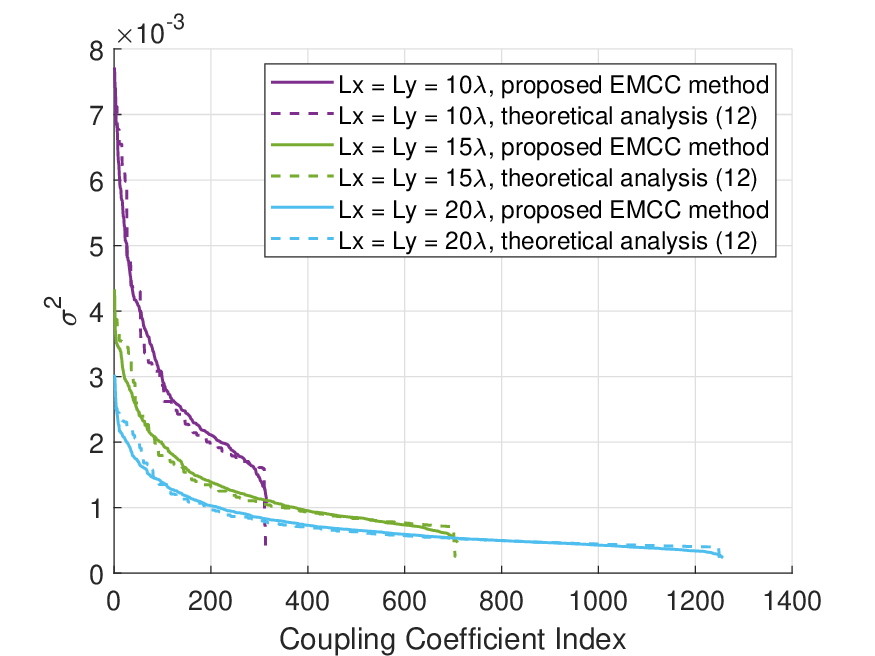}}
\vspace{-3mm}
\caption{Coupling Coefficients of different sizes of antenna array.}
\label{fig::simvscal}
\vspace{-3mm}
\end{figure}


\begin{figure*}[ht]
	\centerline{\includegraphics[width=17cm,height=4.7cm]{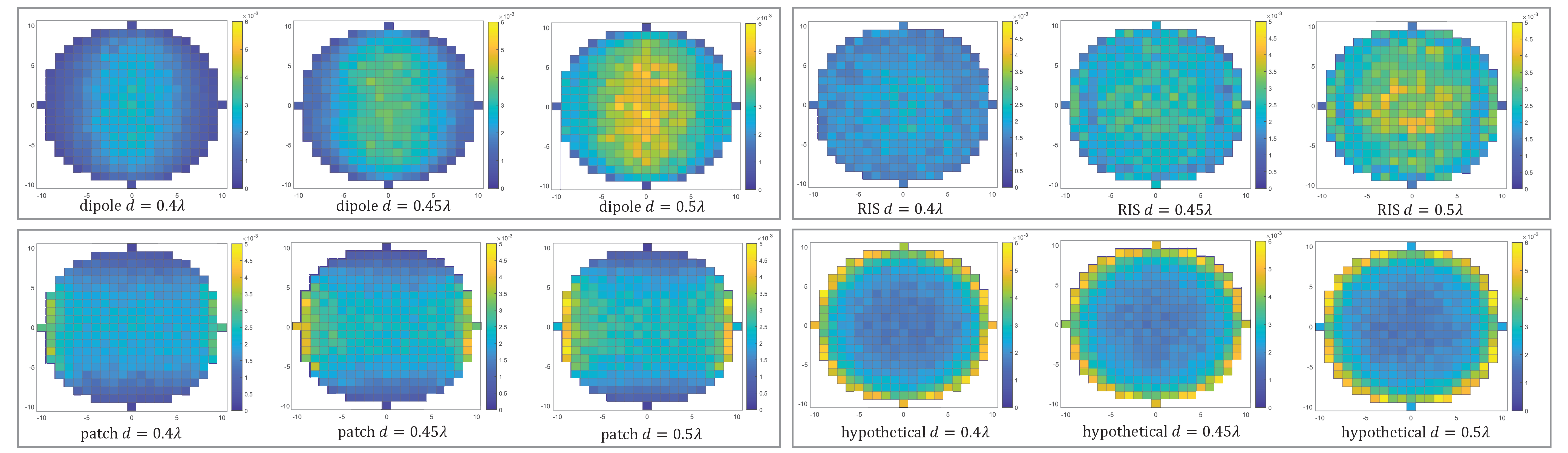}}
	\vspace{-3mm}
	\caption{The distributions of coupling coefficients of different antenna types and element spacing.}
	\label{fig::wndcc}
	\vspace{-3mm}
\end{figure*}

\begin{figure}[t]
	\centerline{\includegraphics[width=5.6cm,height=4cm]{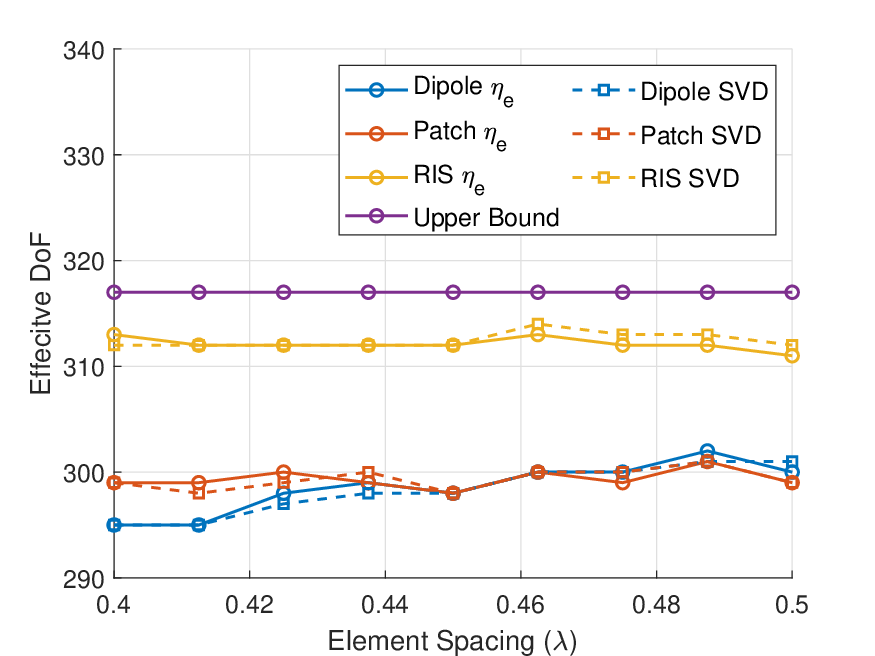}}
	\vspace{-3mm}
	\caption{EDoF vs. element spacing of different antenna types.}
	\label{fig::dof}
	\vspace{-5mm}
\end{figure}

\begin{figure}[t]
	\centerline{\includegraphics[width=5.6cm,height=4cm]{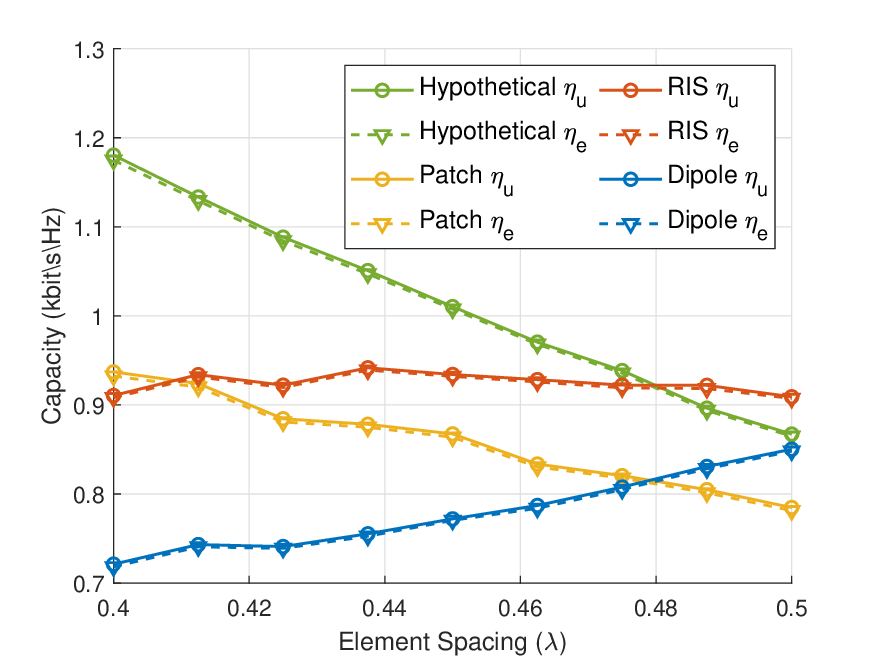}}
	\vspace{-3mm}
	\caption{Ergodic channel capacity vs. element spacing of different antenna types.}
	\label{fig::capacity}
	\vspace{-6mm}
\end{figure}

Fig.\ref{fig::simvscal} shows the coupling coefficients ${\sigma}_T^2$ for different sizes of antenna arrays. To verify the proposed EMCC method, the hypothetical antenna element with $G(\theta,\phi) = 1$ is applied so that ${\sigma}_T^2$ is obtained via both theoretical analysis in (\ref{equ::calculation_cc}) and EMCC for comparison.  
First, ${\sigma}_T^2$ obtained by two methods match each other, which validates the proposed EMCC method. Besides, ${\sigma}_T^2$ decreases as the array size scales up. This is because the wavenumber domain region {\color{blue}$K_T(m_x,m_y)$} that is integrated over in (\ref{equ::calculation_cc}) to obtain ${\sigma}_T^2(m_x,m_y)$ is inversely proportional to the aperture size.

Fig.~\ref{fig::wndcc} illustrates the distribution of coupling coefficients in the wavenumber domain for various antenna types and element spacing. Distinct distributions are demonstrated because of the difference in radiation patterns. Besides, the increase of element spacing $d$ drives the growth of coupling coefficients for  RIS, patch, and dipole antenna array because the realized gain $G$ increases and the {\color{blue} mutual} coupling effect between antenna elements weakens. Moreover, since the mutual coupling effect is omitted for the hypothetical antenna element, its distribution coupling coefficients is irrelevant to element spacing.

Fig.~\ref{fig::dof} shows the EDoF vs. element spacing $d$ for different types of antennas. The solid curves are obtained with (\ref{equ::wndof}) and (\ref{equ::wndof2}) using the statistical information of the wavenumber-domain channel. The dotted curves are deterministic results via performing SVD to the MIMO channel and counting the number of dominant singular values\cite{dofcount}. 
As illustrated by Fig.~\ref{fig::dof}, RIS reach a larger EDoF than patch or dipole antenna array with an increased $d$, because distribution of coupling coefficients of RIS is more even than patch and dipole, as shown in Fig. \ref{fig::wndcc}.
Besides, it can be observed that EDoF calculated based on the statistical information of the channel is consistent with the deterministic results, indicating that the proposed method to calculate the EDoF can describe the number of communication modes accurately.

Fig.~\ref{fig::capacity} shows the ergodic channel capacity vs. element spacing based on (\ref{equ::channelcapwnd}) for different types of antennas.
 The relationship between ergodic channel capacity and the element spacing varies with antenna type. For the antenna array with hypothetical antenna element and the patch antenna array, the decrease in element spacing leads to an increase in channel capacity. For the dipole antenna array, the maximum ergodic channel capacity is reached when element spacing is 0.5$\lambda$. For RIS, $0.4375 \lambda$ is the optimal element spacing to reach the maximum ergodic channel capacity, which is close to the element design in \cite{RIS}, where a $0.434 \lambda$ element spacing is adopted. Moreover, the dotted curves and the solid curves are in good agreement, which indicates that the number of communication modes with effective data transmitting ability matches EDoF well. 


\vspace{-5mm}
\section{Conclusion}\label{sec::conclusion}
In this letter, considering the existence of antenna directivity, we investigated the EDoF and the channel capacity of the XL-MIMO via the analysis of directivity-aware coupling coefficients in the wavenumber domain, given the isotropic scattering Rayleigh-fading channel. 
 Based on the theoretical analysis and simulation results, we drew the following conclusions. 
{\color{black}\emph{First,} theoretical EDoF calculated based on the directivity-aware coupling coefficients, which revealed the statistical information of wavenumber-domain channel, matched well with the simulated EDoF result. 
\emph{Second,} an even distribution of coupling coefficients led to a large EDoF. A RIS could achieve a higher EDoF than the considered dipole and patch antenna arrays when RIS had an evener coupling coefficient distribution.
\emph{Third,} both the antenna type and element spacing affected the antenna directivity, leading to the change of channel capacity.} 
\vspace{-4mm}

\end{document}